\documentclass{article}
\usepackage{spconf}
\usepackage{amssymb}
\usepackage[T1]{fontenc}
\usepackage{placeins}
\usepackage{float}
\usepackage{ragged2e}
\usepackage{fancyhdr}
\usepackage{booktabs}
\usepackage{graphicx}
\usepackage{amssymb,amsmath,bm}
\usepackage{textcomp}
\usepackage{graphicx}
\usepackage{wrapfig}
\usepackage{lipsum}
\usepackage{dblfloatfix}
\usepackage{fixltx2e}
\usepackage{nameref}
\usepackage{amsmath}
\usepackage{graphicx}
\usepackage{varioref}
\usepackage{booktabs}  
\usepackage{siunitx}
\usepackage{numprint}
\usepackage{color}
\usepackage{textcomp}

\usepackage{lipsum}
\begin{document}
%
\name{Debanjan Borthakur\textsuperscript{1}, Harishchandra Dubey\textsuperscript{2}\thanks{\textcolor{blue}{This material is presented to ensure timely dissemination of scholarly and technical work. Copyright and all rights therein are retained by the authors or by the respective copyright holders. The original citation of this paper is:
			Debanjan Borthakur, Harishchandra Dubey, Nicholas Constant, Leslie Mahler, Kunal Mankodiya, Smart Fog: Fog Computing Framework for Unsupervised Clustering Analytics in Wearable Internet of Things, 5th IEEE Global Conference on Signal and Information Processing GlobalSIP 2017, November 14-16, 2017, Montreal, Canada.}}, Nicholas Constant\textsuperscript{1}, Leslie Mahler\textsuperscript{3}, Kunal Mankodiya\textsuperscript{1}\thanks{The research discussed in this manuscript was supported by National Institute of Health Grant:
R01MH108641.}
}

\address{\textsuperscript{1}Wearable Biosensing Lab, University of Rhode Island, Kingston, RI-02881, USA\\
\textsuperscript{2} The University of Texas at Dallas, Richardson, TX-75080, USA\\
\textsuperscript{3}Department of Communicative Disorders, University of Rhode Island, Kingston, RI-02881, USA\\ 
Email: (kunalm@uri.edu, lmahler@uri.edu)
}
\title{Smart Fog: Fog Computing Framework for Unsupervised Clustering Analytics in Wearable Internet of Things}
\maketitle
\begin{abstract}
The increasing use of wearables in smart telehealth generates heterogeneous medical big data. Cloud and fog services process these data for assisting clinical procedures. IoT based e-healthcare have greatly benefited from efficient data processing. This paper proposed and evaluated use of low-resource machine learning on Fog devices kept close to the wearables for smart healthcare. In state-of-the-art telecare systems, the signal processing and machine learning modules are deployed in the cloud for processing physiological data. We developed a prototype of Fog-based unsupervised machine learning big data analysis for discovering patterns in physiological data. We employed Intel Edison and Raspberry Pi as Fog computer in proposed architecture. We performed validation studies on real-world pathological speech data from in-home monitoring of patients with Parkinson's disease (PD). Proposed architecture employed machine learning for analysis of pathological speech data obtained from smartwatches worn by the patients with PD. Results showed that proposed architecture is promising for low-resource clinical machine learning. It could be useful for other applications within wearable IoT for smart telehealth scenarios by translating machine learning approaches from the cloud backend to edge computing devices such as Fog.
\end{abstract}
\maketitle
\begin{figure*}[!t]
\centering
\includegraphics[width=480pt]{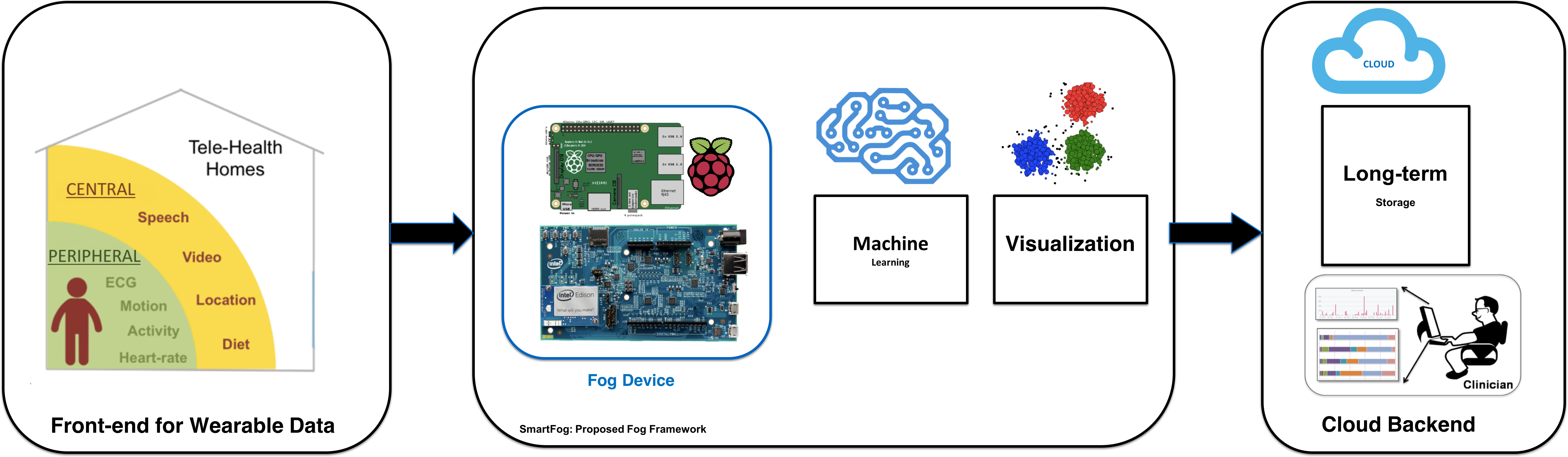}
\vspace{-2mm}
\caption{Proposed SmartFog architecture for enhanced analytics in wearable internet of medical things. It is developed and evaluated for telehealth application.}
\label{fig_sfog}
\end{figure*}
\section{Introduction}
As described in~\cite{chiang2016fog} Fog is a new architecture for computing, storage, control and networking that brings these services closer to end users.In simple words, the decentralization of services at the edge of the network is achieved.The computation and control closer to the sensors make the concept of Fog a better alternative to the cloud.In our proposed architecture of smart Fog, we leveraged the idea of Fog for speech signal processing for telehealth monitoring. Speech signal processing and Machine learning are fundamental blocks for detection and evaluation of speech disorders like dysarthria in patients with Parkinson's diseases that affects a significant portion of the world population. Telehealth monitoring is very
effective for the speech-language pathology, and smart devices
like EchoWear~\cite{r1_dubey2015echowear} can be useful in such situations. Several signs indicate the relationship of dysarthria, speech
prosody, and acoustic features. As authors in~\cite{r2_zhao2014automatic} mentions dysarthria always accompanies patients with Parkinson's disease
Characterized by the monotony of speech, reduced stress, variable rate, imprecise consonants, and a breathy and harsh voice Authors in~\cite{r3_falk2012characterization}~\cite{r4_patel2012relationship} suggested that extreme F0 variation and range in speakers with severe dysarthria exist. Another important acoustic feature for dysarthria is the amplitude of the speech uttered by the patients with Parkinson's disease. In~\cite{r5_watts2016retrospective} authors mention about reduced vocal intensity in hypokinetic dysarthria in Parkinson disease.
This paper presents a Fog Computing architecture, SmartFog that relied on unsupervised clustering for discovering patterns in pathological speech data obtained from patients with Parkinson's disease(PD).
The patients with PD use smartwatch while performing speech exercises at home. The speech data were routed into the Fog computer via a nearby tablet/smartphone. The Fog computer extracts loudness and fundamental frequency features for quantifying pathological speech. The speech features were normalized and processed with k-means clustering. When we see an abnormal change in features,  results are uploaded to the cloud. In other situations, data is only processed locally. In this way, Fog device could perform "smart" decision on when to upload the data to cloud backend and when not. We developed two prototypes using Intel Edison and Raspberry Pi. Both of the prototypes were used for comparative analysis of computation time. Both systems were tested on real world pathological speech data from telemonitoring of patients with Parkinson's disease.
The increasing use of wearables in smart telehealth system led to the generation of huge medical big data~\cite{dubey2017fog,dubey2016harmonic,dubey2016bigear,dubey2015multi}. 
The telehealth services leverage these data for assisting clinical procedures. This paper suggests use of low-resource machine learning on Fog devices kept close to the wearable for smart telehealth. For traditional telecare systems, the signal processing and machine learning modules are deployed in the cloud that processes physiological data.
In our analysis, we have chosen the average fundamental frequency(F0) in hertz and
average intensity in decibel for K-means clustering
analysis. The algorithm efficiently clusters the unlabeled data
into groups of similarity that was done on the fog platform.One use of this analysis can be for real time  Parkinson's phenotypic sub-groupings based on the clusters.
\begin{figure*}[ht!] 
\centering
\includegraphics[width=480pt,trim={10 18 10 18},clip]{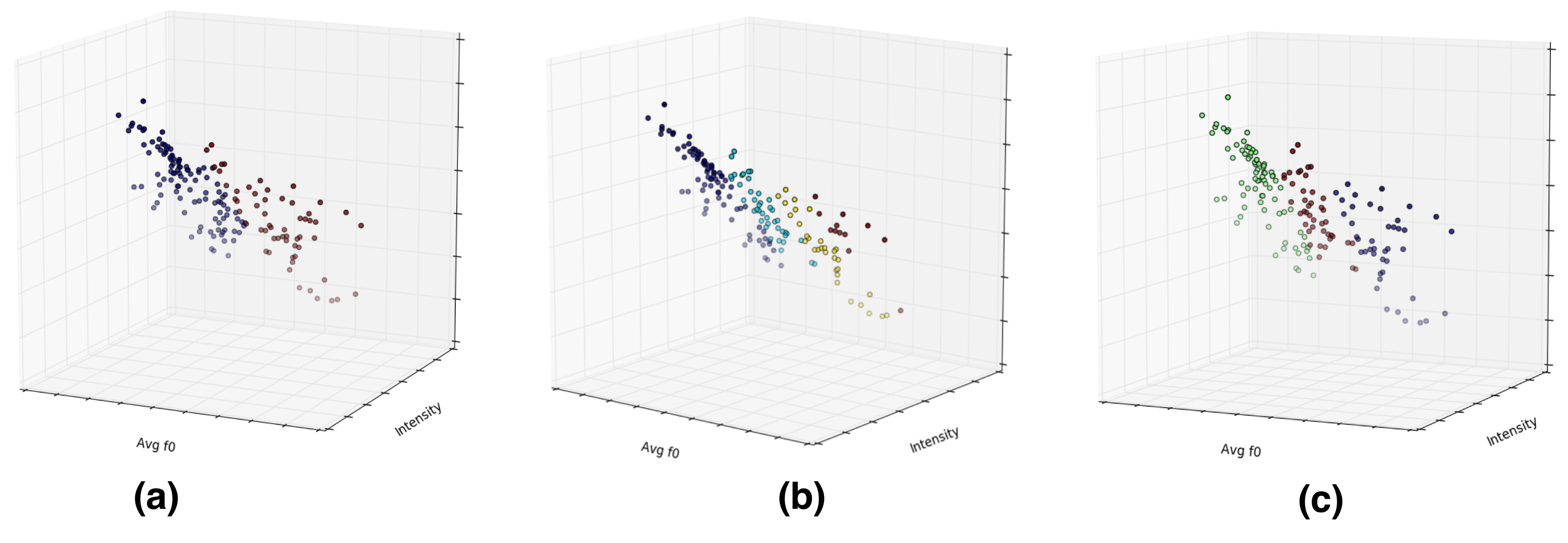}
\caption{K-means clustering plot}
\end{figure*}
\vspace{-2mm}
\section{Related Works} 
\subsection{Telehealth and Associate Challenges}
The Fog Architecture shifts computation, networking, and storage to the edge of the network.Various authors have described a different architecture for Fog. FIT as described in~\cite{r6_monteiro2016fit} has the following components: (1) Smartwatch; (2) Fog computer; and (3) Cloud backend. In the paper~\cite{r11_hiremath2014wearable} authors presents an effort to conceptualize WIoT concerning their design, function, and applications.The paper~\cite{r10_dubey2015fog} demonstrates the Fog Data that is a service oriented architecture for Fog computing.This literature emphasizes the importance and versatility of Fog computing.The challenges IoT faces are described in~\cite{chiang2016fog} are the requirement of stringent low latency, IoT applications such as gaming, virtual reality demands this. The issue of Network Bandwidth and Resource-constrained devices are another challenges to the emerging field of IoT. Thus arises the importance of fog that distributes computing, control, storage and networking functions closer to the end user~\cite{chiang2016fog}.
\subsection{Big data and Telehealth}
Tele-Health utilizes the recent developments of Big Data in the context of biomedical and healthcare.Fields like medical and health informatics, translational bioinformatics, sensor informatics etc can avail the benefit of the personalized information from a diverse range of data sources\cite{andreu2015big,barik2018fog}. Authors in \cite{r10_dubey2015fog}, proposes, validates and evaluates Fog Data architecture for Fog computing. The proposed architecture is a low power embedded computer that carries out data mining and  analysis on data collected from various wearable sensors used for telehealth applications.\cite{cancela2013telehealth} mentions about European project 'PERFORM' that is a sophisticated multi-parametric system FOR the continuous effective assessment and monitoring of motor status in Parkinson Disease and other neurodegenerative diseases. It provides a telehealth system for remote monitoring of Parkinson Patients.The paper also summarizes the technical performance of the system and the feedback received from the patients in terms of usability and wearability.We in our work used Parkinson speech data analysis for our proposed smart-fog framework.

\subsection{Wearable Internet of Thing for Telehealth}
IoT Device that interacts with the fog node is composed of sensors that are capable of collecting and transmitting data via wireless means.IoT allows handling of objects remotely across the network. The versatility of IoT makes it more suitable for smart grids, smart homes, smart cities and wearable health monitoring systems.This paper focuses on the health aspect of IoT.Integration with the internet offers IoT devices an IP address for better communication. Big data and Internet of Things work collectively, and we tried to leverage this relationship in our proposed architecture.
%
We used Raspberry Pi and Intel Edison as Fog computing device for the analysis discussed in this paper. Fog Interface as described in~\cite{r6_monteiro2016fit, constant2017fog} is a low-power embedded computer that acts as a smart interface between the smartwatch and the cloud. It is used for collection, storage, and processing of the data before sending features to secure cloud storage.  Raspberry Pi is used as Fog device for this work. The Raspberry Pi is a series of credit card-sized single board computers that has gained much popularity owing to its small size and multipurpose utility. It has ARM compatible central processing unit and on-chip graphic processing units.
\subsection{Fog Computing}
Cloud computing provides shared computer processing and data analysis, in other terms Cloud is a hub of computing resources such as computer networks, servers, storage, and services. The availability of high-capacity networks, low-cost computers, and storage devices makes cloud a highly demanded service for the users seeking for high computing power. Cloud can interact with the IoT device via the fog node. This paper concentrates on the side of fog computing, which allows users a higher computing power at the instrument end.Reliance on fog will help cut the costs associated with the Cloud to an extent~\cite{barik2017mistgis,barik2017foggeo}.
\vspace{-2mm}
\section{Fog-based Low-Resource Machine Learning}
\vspace{-2mm}
\subsection{Feature Extraction}
Feature engineering is the initial step in any machine learning analysis. It is the process of proper selection of data metric to input as features into a machine learning algorithm. In K-means clustering analysis, the selection of features that are capable of capturing the variability of the data is essential for the algorithm to find the groups based on similarity. Our subjects were patients with Parkinson's disease and the features chosen were the average fundamental frequency (F0) and Average amplitude of the speech utterance. Speech data from the patients with Parkinson's disease were collected. For analysis, 164 speech samples were considered.These samples comprised of sound files with utterances as a short /a/, a long /a/, a normal then high pitched /a/, a normal then low pitched /a/ and phrases.
The feature extraction is done with the help of Praat scripting language~\cite{r7_boersma2002praat}. For pitch , the algorithm performs an acoustic periodicity detection on the basis of an accurate autocorrelation method.  For calculating the intensity the values in the sound are first squared, then convolved with a Gaussian analysis window. The intensity is calculated in decibels.
\subsection{K-means Clustering}
K-means clustering is a type of unsupervised learning, that is used for exploratory data analysis of no labeled data~\cite{bishop1995neural}. K-means is a method of vector quantization and is quite extensively used in data mining.The goal of this algorithm is to find groups in the data, the number of groups represented by the variable K. The algorithm works iteratively to assign each data point to one
of K groups based on the features that are provided. The input to the algorithm are the features, and the value of K.  K centroids are initially randomly selected, then the algorithm iterates until convergence. This algorithm aims to minimize the squared error function J.
\[
J=\sum^K_{k=1} \sum_{i\in c_k} || x_i - m_k ||^2 \ \ \ \ \ \ \ \rm 
\]
Where  Euclidean distance is chosen between the data point and cluster center. Feature Engineering is an essential part of this algorithm. Authors in~\cite{r8_kadhim2015novel}, uses optimized K-means, that clusters the statistical properties such as the variance of the probability density functions of the clusters extracted features. In~\cite{r9_majeed2012hierarchical} the authors have used clustering on a database containing feature vectors derived from Malay digits utterances. The features extracted in~\cite{r9_majeed2012hierarchical} were the Mel-Frequency Cepstral Coefficients (MFCC). In our work, we have chosen the average fundamental frequency and average intensity as features extracted from the speech files for applying K-means clustering.
\section{Results \& Discussions}
For our analysis we have chosen speakers with 164 speech samples with utterances that are a short /a/, a long /a/, a normal then high pitched /a/,a normal then low pitched /a/ and phrases. The features were chosen are average fundamental frequency and intensity. Feature extraction is done using praat~\cite{r7_boersma2002praat} an acoustic analysis software and using Praat scripts that use standard algorithms to extract pitch and intensity mentioned in the discussion above. The results are shown in the form of plots.The k-means clustering analysis is done on Python programming language.The plots below show
the Clusters of the speech data samples used in the analysis.Different colors represent different mutually exclusive groups. The analysis is done with 2, 3 and 4 number of clusters, i.e. the value of k chosen as 2 and 3 and four respectively.\\
Figure (a) shows the K-means clustering plot for 2 clusters shown with different colors.The python script is run on Raspberry Pi and Intel Edison to generate the results.\\
Figure (b) displays the k-means cluster plot for 4
Clusters designated with four different colors in a 3D plot.Each observation belongs to the cluster with the nearest mean in k-means clustering.We have used k-means for feature learning performed in the fog device.\\
Figure(c) shows the k-means clustering plot for 3 clusters with different colors in 3D.The python script run on Raspberry Pi and Intel Edison was used for generating the results displayed in the figure.
\subsection{Performance Comparison}
The Raspberry Pi provides a low-cost computing terminal. The Edison  is  a deeply embedded IoT computing module.   There is a difference of processor speed and power consumption in Edison and Raspberry Pi . The Machine Learning algorithms were run on both of the devices and their Run time, average CPU usage and Memory usage have been calculated.
\begin{figure}[!t]
\centering
\includegraphics[width=260pt,trim=4 4 4 4,clip]{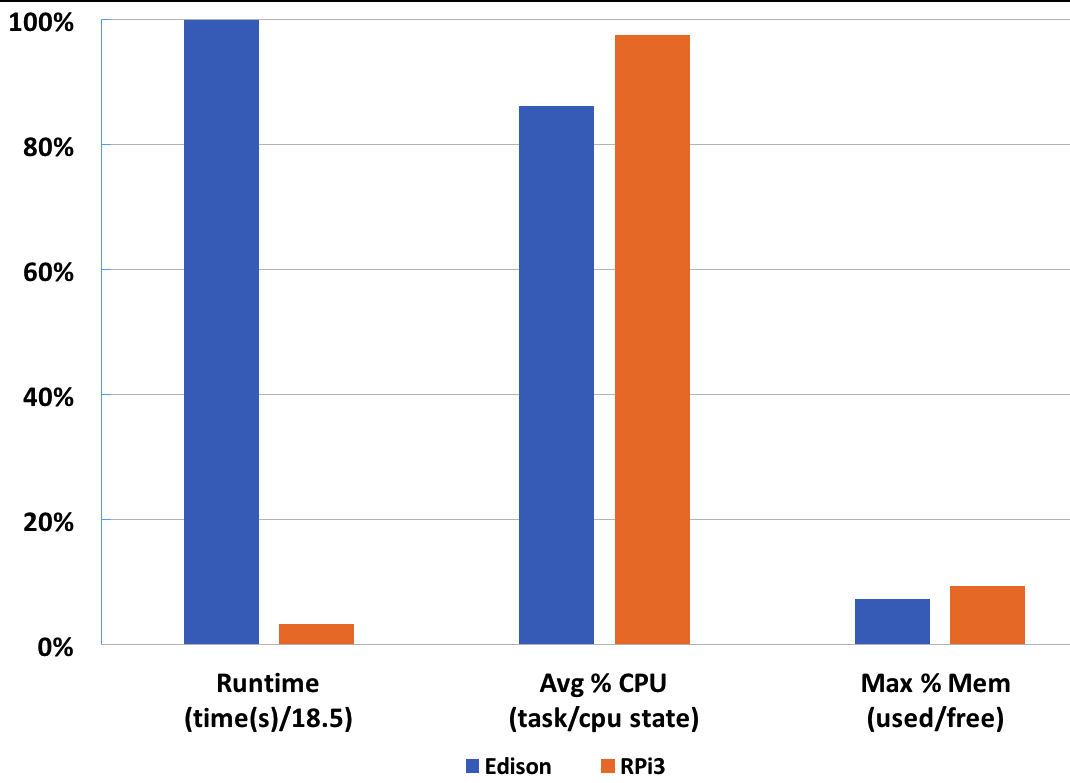}
\caption{A comparison of Intel Edison and Raspberry pi.}
\end{figure}
Figure 3 shows the comparison of Intel Edison and raspberry Pi fog devices.The ideal system will minimize runtime, maximize CPU usage, and use a modest amount of memory. The raspberry Pi either outperformed or matched the Edison in each of this criterion. The raspberry Pi was not capable of generating a graphical output for this type of analysis in a real-time response threshold of 200ms. However, without a need for complex graphics, the raspberry Pi was able to reach the threshold clocking in at 160ms.
\section{Conclusions}
Fog computing emphasizes proximity to end-users unlike cloud computing along with  local resource pooling, reduction in latency, better quality of service and better user experiences. This paper relied on Fog computer for low-resource machine learning. As a use case, we employed K-means clustering on clinical speech data obtained from patients with Parkinson's disease (PD).Proposed Smart-Fog
architecture can be useful for health problems
like speech disorders and clinical speech processing in real
time as discussed in this paper.Fog computing reduced the
onus of dependence on Cloud services with availability of big data.There will be more
aspects of this proposed architecture that can be investigated
in future.We can expect Fog architecture to be crucial in
shaping the way big data handling and processing happens
in near future. 
%
\section{Acknowledgement}
Authors would like to thank George and  Anne Ryan Institute for Neuroscience for their support and help.
\bibliographystyle{IEEEbib}
\bibliography{ref}

\begin{thebibliography}{10}

\bibitem{chiang2016fog}
M.~Chiang and T.~Zhang,
\newblock ``Fog and iot: An overview of research opportunities,''
\newblock {\em IEEE Internet of Things Journal}, vol. 3, no. 6, pp. 854--864,
  2016.

\bibitem{r1_dubey2015echowear}
H.~Dubey, J.~C Goldberg, M.~Abtahi, L.~Mahler, and K.~Mankodiya,
\newblock ``{EchoWear:} smartwatch technology for voice and speech treatments
  of patients with parkinson's disease,''
\newblock in {\em Proceedings of the conference Wireless Health}. ACM, 2015.

\bibitem{r2_zhao2014automatic}
S.~Zhao, F.~Rudzicz, L.~G. Carvalho, C.~M{\'a}rquez-Chin, and S.~Livingstone,
\newblock ``Automatic detection of expressed emotion in parkinson's disease,''
\newblock in {\em IEEE ICASSP}, 2014.

\bibitem{r3_falk2012characterization}
T.~H. Falk, W.~Chan, and F.~Shein,
\newblock ``Characterization of atypical vocal source excitation, temporal
  dynamics and prosody for objective measurement of dysarthric word
  intelligibility,''
\newblock {\em Speech Communication}, vol. 54, no. 5, pp. 622--631, 2012.

\bibitem{r4_patel2012relationship}
R.~Patel, K.~C. Hustad, K.~P. Connaghan, and W.~Furr,
\newblock ``Relationship between prosody and intelligibility in children with
  dysarthria,''
\newblock {\em Journal of medical speech-language pathology}, vol. 20, no. 4,
  2012.

\bibitem{r5_watts2016retrospective}
C.~R. Watts,
\newblock ``A retrospective study of long-term treatment outcomes for reduced
  vocal intensity in hypokinetic dysarthria,''
\newblock {\em BMC Ear, Nose and Throat Disorders}, vol. 16, no. 1, pp. 2,
  2016.

\bibitem{dubey2017fog}
H.~Dubey, N.~Constant, A.~Monteiro, M.~Abtahi, D.~Borthakur, L.~Mahler, Y.~Sun,
  Q.~Yang, and K.~Mankodiya,
\newblock ``Fog computing in medical internet-of-things: Architecture,
  implementation, and applications,''
\newblock in {\em Handbook of Large-Scale Distributed Computing in Smart
  Healthcare}. 2017, Springer International Publishing AG.

\bibitem{dubey2016harmonic}
H.~Dubey, R.~Kumaresan, and K.~Mankodiya,
\newblock ``Harmonic sum-based method for heart rate estimation using ppg
  signals affected with motion artifacts,''
\newblock {\em Journal of Ambient Intelligence and Humanized Computing}, pp.
  1--14, 2016.

\bibitem{dubey2016bigear}
H.~Dubey, M.~R. Mehl, and K.~Mankodiya,
\newblock ``Bigear: Inferring the ambient and emotional correlates from
  smartphone-based acoustic big data,''
\newblock in {\em EEE First International Conference on Connected Health:
  Applications, Systems and Engineering Technologies (CHASE), Washington, DC,
  2016, pp. 78-83.}, 2016, number doi: 10.1109/CHASE.2016.46.

\bibitem{dubey2015multi}
H.~Dubey, J.~C. Goldberg, K.~Mankodiya, and L.~Mahler,
\newblock ``A multi-smartwatch system for assessing speech characteristics of
  people with dysarthria in group settings,''
\newblock in {\em IEEE 16th International Conference on e-Health Networking,
  Applications and Services (Healthcom),}, 2015.

\bibitem{r6_monteiro2016fit}
A.~Monteiro, H.~Dubey, L.~Mahler, Q.~Yang, and K.~Mankodiya,
\newblock ``Fit: A fog computing device for speech tele-treatments,''
\newblock in {\em IEEE Smart Computing (SMARTCOMP)}, 2016.

\bibitem{r11_hiremath2014wearable}
S.~Hiremath and K.~Yang, G.and~Mankodiya,
\newblock ``Wearable internet of things: Concept, architectural components and
  promises for person-centered healthcare,''
\newblock in {\em Mobihealth Conference}. IEEE, 2014.

\bibitem{r10_dubey2015fog}
H.~Dubey, J.~Yang, N.~Constant, A.~M. Amiri, Q.~Yang, and K.~Makodiya,
\newblock ``Fog data: Enhancing telehealth big data through fog computing,''
\newblock in {\em Fifth ASE BigData 2015, Kaohsiung, Taiwan}. ACM.

\bibitem{andreu2015big}
J.~Andreu-Perez, C.~C.~Y. Poon, R.~D. Merrifield, S.~T.C. Wong, and G.~Yang,
\newblock ``Big data for health,''
\newblock {\em IEEE journal of biomedical and health informatics}, vol. 19, no.
  4, pp. 1193--1208, 2015.

\bibitem{barik2018fog}
R.~K. Barik, H.~Dubey, C.~Misra, D.~Borthakur, N.~Constant, S.~A. Sasane, R.~K.
  Lenka, B.~S.~P. Mishra, H.~Das, and K.~Mankodiya,
\newblock ``Fog assisted cloud computing in era of big data and
  internet-of-things: Systems, architectures and applications,''
\newblock in {\em Cloud Computing for Optimization: Foundations, Applications,
  Challenges}, p.~23. Springer, 2018.

\bibitem{cancela2013telehealth}
J.~Cancela, M.~Pastorino, M.~T. Arredondo, and O.~Hurtado,
\newblock ``A telehealth system for parkinson's disease remote monitoring. the
  perform approach,''
\newblock in {\em 35th Annual International Conference of the IEEE Engineering
  in Medicine and Biology Society (EMBC),}, 2013, pp. 7492--7495.

\bibitem{constant2017fog}
N.~Constant, D.~Borthakur, M.~Abtahi, H.~Dubey, and K.~Mankodiya,
\newblock ``Fog-assisted wiot: A smart fog gateway for end-to-end analytics in
  wearable internet of things,''
\newblock in {\em The 23rd IEEE Symposium on High Performance Computer
  Architecture HPCA 2017,Austin, Texas, USA}, 2017.

\bibitem{barik2017mistgis}
R.~Barik, H.~Dubey, R.~K. Lenka, K.~Mankodiya, T.~Pratik, and S.~Sharma,
\newblock ``Mistgis: Optimizing geospatial data analysis using mist
  computing,''
\newblock in {\em International Conference on Computing Analytics and
  Networking (ICCAN 2017)}. Springer, 2017.

\bibitem{barik2017foggeo}
R.~K. Barik, H.~Dubey, R.~K. Lenka, N.V.R. Simha, S.~A. Sasane, C.~Misra, and
  K.~Mankodiya,
\newblock ``Fog computing-based enhanced geohealth big data analysis,''
\newblock in {\em 2017 International Conference on Intelligent Computing and
  Control (I2C2)}. IEEE, 2017.

\bibitem{r7_boersma2002praat}
Paulus Petrus~Gerardus Boersma et~al.,
\newblock ``Praat, a system for doing phonetics by computer,''
\newblock {\em Glot international}, vol. 5, 2002.

\bibitem{bishop1995neural}
Christopher~M Bishop,
\newblock {\em Neural networks for pattern recognition},
\newblock Oxford university press, 1995.

\bibitem{r8_kadhim2015novel}
H.~A. Kadhim, L.~Woo, and S.~Dlay,
\newblock ``Novel algorithm for speech segregation by optimized k-means of
  statistical properties of clustered features,''
\newblock in {\em IEEE Progress in Informatics and Computing (PIC) Conference},
  2015.

\bibitem{r9_majeed2012hierarchical}
S~Majeed, H~Husain, S~Samad, and A~Hussain,
\newblock ``Hierarchical k-means algorithm applied on isolated malay digit
  speech recognition,''
\newblock {\em International proceedings of computer science \& information
  technology}, vol. 34, pp. 33--37, 2012.

\end{thebibliography}
\end{document}